# Persistent photoconductivity in oxygen deficient $YBa_2Cu_3O_{7-\delta}$/ $La_{2/3}Ca_{1/3}MnO_{3-x}$ superlattices grown by Pulsed Laser Deposition

Kazuhiro. Kawashima[a], Soltan. Soltan[a,b], Gennady. Logvenov[a], and Hanns-Ulrich Habermeier[a*]

a.) *Max Planck Institute for Solid State Research, Heisenbergstr 1, D 70569 Stuttgart, Germany*

b.) *Department of Physics, Faculty of Science, Helwan Unversity, 11798-Helwan Cairo, Egypt.*

**Abstract:**

We report a large persistent photoconductivity (PPC) in oxygen-reduced $YBa_2Cu_3O_{7-\delta}$/ $La_{2/3}Ca_{1/3}MnO_{3-y}$ superlattices grown by pulsed laser deposition that scales with oxygen deficiency and is similar to that observed in single layer $YBa_2Cu_3O_{7-\delta}$ films. These results contradict previous observations, where, in sputtered bilayer samples, only a transient photoconductivity was found. We argue that the PPC effect in superlattices is caused by the PPC effect due to $YBa_2Cu_3O_{7-\delta}$ layers with limited charge transfer to $La_{2/3}Ca_{1/3}MnO_{3-y}$. The discrepancy arises from the different permeability of charges across the interface and sheds light on the sensitivity of oxide interface properties to details of their preparation.

[*]corresponding author: Hanns-Ulrich Habermeier, huh@fkf.mpg.de

Recent advances in the synthesis of epitaxial high quality complex oxide heterostructures and superlattices open an avenue for the exploration and manipulation of the electronic phase behavior of artificially layered structures.[1-3] Oxide heterostructures and superlattices combining materials with different collective electronic ordering phenomena especially increased in popularity due to the possibility to explore the interactions between the long-range order of electrons and the modification of their spin and orbital states by artificial architectures.[4-7] They offer an unique opportunity to externally control their electrical, magnetic and optical properties based on the coupled charge, spin, orbital, and lattice interactions of the constituents by exposing them either to elastic, electrical or magnetic fields or to subject them to a pulsed or continuous-wave photon flux. The physics behind is seen in the subtle balance of a rich set of coexisting electronic phases with comparable ground state energies. Advanced architectures can be constructed, that show functions well beyond the charge density manipulations that determine the functionality of conventional semiconductor heterostructures. Interface engineering, currently a flourishing field, enables the tuning of the properties of such heterostructures and might pave the way to access different quantum phases. To explore these opportunities, a fundamental understanding of the modifications of the electronic structure at the interface is required. Photon exposure of complex oxides is one possibility to achieve this goal, as it alters the charge carrier density and thus changes the interplay between different phases. In particular, samples with oxygen vacancies can lead to photogenerated effects based on photodoping and may direct to so far unknown device applications. Reciprocally, photodoping can be used as an analytical tool to detect oxygen vacancies in such systems. Tailored changes of the electronic properties of complex oxides by intrinsic photoinduced doping have been observed in many perovskite-type materials, amongst them the superconducting cuprates[8,9] and the colossal magnetoresistance manganites are the most prominent examples.[10,11] The majority of these contributions cover work using thin films. Here, however, for films with a thickness smaller than or comparable to the optical penetration depth (typically < 100 nm for semimetallic manganites and $YBa_2Cu_3O_{7-\delta}$ (YBCO) as well), the influence due to charge injection from the substrate and/or the substrate/film interface has not been to be considered with a few exceptions.[12-14]

Photoexcitation in underdoped cuprates, especially YBCO, shows enhancement of superconductivity as well as both, transient (TPC) and persistent (PPC) photoconductivity and is most pronounced in the UV part of the spectrum.[15] The photoconductivity is intrinsic to the material and is observable in single crystals and thin films deposited onto various types of substrates (e.g. MgO, $SrTiO_3$). Two different mechanisms can account for the PPC in YBCO. One is based on the generation of excess holes by trapping the electrons of photoexcited electron-hole pairs in oxygen vacancies of the Cu-O chains with the remaining hole being transferred to the $CuO_2$ plane.[8,9,16] Alternatively, the photoassisted chain-ordering model assumes that the light exposure induces an oxygen reordering process giving rise to an increase of the average length of the Cu-O-chain segments. This in turn results in an increase of the hole concentration in the CuO$_2$ plane.[17]

In the case of manganites, the investigation of photoinduced phenomena is a relatively recent field and a plethora of photoinduced effects has been observed, ranging from an X-ray induced metal-insulator transition combined with a structural change in charge-ordered Pr$_{0.7}$Ca$_{0.3}$MnO$_3$ single crystals[18] to a reversible photoinduced switching between a charge ordered (CO) - orbital ordered insulating (OOI) and a ferromagnetic metallic (FMM) state in Pr$_{0.55}$(Ca$_{1-y}$Sr$_y$)$_{0.45}$MnO$_3$ single crystals.[19] Although, no general understanding of the microscopic mechanisms of photoinduced phenomena in manganites is currently available, three main mechanisms should be distinguished as comprehensively reviewed by Beyreuther *et al*.[13] First, in materials with a composition close to a bicritical region between the CO/OOI and the FMM phases photon induced excitations across the charge gap are the predominant mechanism for phase changes. Second, photoexcitation with photon energies above the polaron binding energy leads to a delocalization of charge carriers. The polaron binding energy is connected to the strong Jahn-Teller effect of the Mn$^{3+}$ ions potentially giving rise to a long-range polaron order which can cause under illumination a collective charge delocalization and a transition into a metastable metallic phase. Third, as shown by Gilabert *et al.,*[10] a PPC effect of less than 10% of the resistivity could be observed at low temperatures (T < 30 K) in the metallic phase of oxygen deficient LaCaBaMnO and Pr$_{2/3}$Sr$_{1/3}$MnO$_3$ thin films upon illumination of UV and/or visible light.

So far, few attempts have been undertaken to use photon exposure as a tool to modify carrier concentrations and mobilities in complex oxide heterostructures and superlattices. Peña *et al.*[20] reported photogenerated effects in oxygen depleted sputtered YBa$_2$Cu$_3$O$_{7-\delta}$/La$_{2/3}$Ca$_{1/3}$MnO$_{3-y}$ (YBCO/LCMO) bilayers and found a transient photoconductivity with a relaxation time 4 orders of magnitude faster than the persistent photoconductivity observed in oxygen depleted single layer high-T$_C$ superconductors of the YBCO family. They discuss their findings in terms of light-induced charge transfer through the interface and enhanced recombination of photogenerated holes in LCMO. Rastogi and Budhani[21] as well as Tebano *et al.*[22] observed a giant photoconductivity in SrTiO$_3$/LaAlO$_3$ heterostructures and explained their results on the same general ground as the PPC effect observed in semiconductor heterostructures.[23,24]

In this paper we investigate photoexcitation effects in carefully oxygen-reduced PLD-grown YBCO/LCMO superlattices (SL's) and demonstrate the existence of a so far unknown persistent photoconductivity. Single YBCO thin films and superlattices consisting of YBCO and LCMO were deposited by conventional pulsed laser deposition (PLD) techniques. For the PLD process, a KrF excimer laser with a wavelength of λ=248 nm was used with the photon fluency adjusted to 1.6 J/cm$^2$, and the pulse frequencies to 2 Hz and 3 Hz for YBCO and LCMO, respectively. All samples were deposited on 5x5 mm$^2$ single crystal (001)-oriented SrTiO$_3$ substrates. The thickness of single layer YBCO samples is fixed to 100 nm, and the LCMO/YBCO SL's are composed of 20 nm of LCMO and 20 nm of YBCO on top with 8 repetitions of them. For the deposition, the substrates were heated to 720 °C in an oxygen partial pressure of 0.2 mbar. All the single YBCO and SL's are grown at this condition. After the deposition, the films are cooled down to 520°C at the ramping rate of 5°C/min, and subsequently the oxygen partial pressure was increased to 1 bar. The samples were annealed for 1 hour to obtain complete oxygenation. To systematically adjust different levels of oxygenation, the samples were then cooled down to 400°C at a rate of 10°C/min, and then exposed to a reduced oxygen atmosphere with the partial pressure ranging from 1 bar to $10^{-3}$ mbar. The annealing times were 2.5 and 16 hours for single YBCO and SL's, respectively. After annealing, the infrared laser heater was turned off and the samples were cooled down to room temperature. To determine the phase purity, orientation and the c-axis lattice parameter of all films, X-ray diffraction was performed at room temperature using a Bruker AXS-D8 X-ray diffractometer with CuK$_\alpha$ radiation (wave length 0.1541 nm).

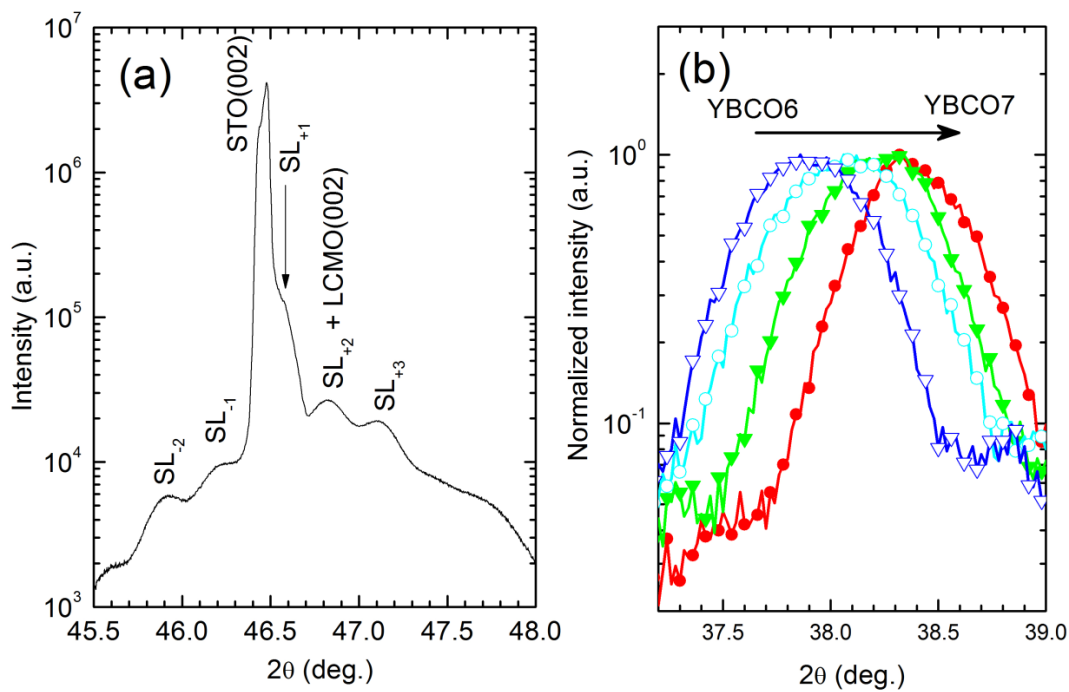

Fig. 1. (a) XRD pattern of a [20nm/20nm]$_8$ YBCO/LCMO superlattice showing the main peaks and superlattice satellites up to the third order. (b) Normalized XRD (005) peak for a [20nm/20nm]$_8$ YBCO/LCMO superlattice with different oxygen concentration.

All single layer YBCO samples showed strong diffraction peaks only from the (00l) family of YBCO. In SL samples, in addition to the YBCO peaks (00l) reflections from LCMO appear together with superlattice satellite peaks, despite of the relatively thick individual layers. In Fig.1a the diffraction pattern around the STO (002) peak is shown for a [20nm/20nm]$_8$ SL, indicating the main peak and the satellites up to the third order. In Fig. 1b the shift of the YBCO (005) peak for a [20nm/20nm]$_8$ YBCO/LCMO superlattice with increasing oxygen depletion is depicted. The quantitative analysis of the data, especially the relationship between the lattice parameter and the oxygen concentration of the SL films, will be published in a forthcoming paper.[25]

For the photoconductivity investigations, the samples were mounted into an optical cryostat (CryoVac –Konti-Cryostat) to measure the transport properties and their photo- induced modifications applying standard four-point probe techniques. First, the resistance was measured in the dark during a heating process from 5K to 270K with a ramping rate of 0.5K/min in the dark. Then, keeping the temperature at 95K, the sample was illuminated by a Xe lamp powered at 600W using a transmission filter (Schott KG3 Glass Filter) to block the infrared part of the spectrum. The effective wave length of the light at the sample site ranges from 350 nm to 600 nm, and the duration time is fixed to 5 hours. During the illumination, the time dependence of resistance, R (t), was measured. After turning off the light, the samples were cooled down to 5 K and the R (T) data were recorded in the heating process. For several representative samples, we carried out further measurements, including the heating process under the illumination after 5-hour illumination. This additional measurement enables us to confirm whether or not our samples show a transient photoconductivity. As a control experiment, we investigated the PPC effect in single layer YBCO films with different degrees of oxygen depletion and confirmed the results of Osquiguil[17] quantitatively. The photo-induced effect in a representative SL sample is given in Fig.2. Fig.2 (a) shows the R (T) data of a strongly oxygen-depleted SL sample in the dark (open triangles), after a 5 h photon exposure at 95K (filled circles) and during illumination after 5h photon exposure (open circles). The $T_C$ of the pristine film shows a value of 22.1 K corresponding to an oxygen concentration of x = 6.5. Fully oxidized films show a $T_C$ of 82 K, a value slightly reduced from the intrinsic value for YBCO caused by the magnetic interaction and orbital reconstruction at the YBCO/LCMO interface.[6,7] We observe a persistent enhancement of $T_C$ by $\Delta T_C = T_C^{after\ ill} - T_C^{dark} = 6.3$ K. The R (T) data for the film after illumination in the dark and during photon exposure coincide in the temperature range 5K < $T_C$ < 200 K indicating vanishing TPC contributions. Above 200 K a transient contribution of ~ 4% is observed, tentatively ascribed to the extended exposure to light or a transient structural reconfiguration of the oxygen ion positions. Fig. 2b shows the time dependence of resistance during illumination at 95K with a stretched exponential behavior of the Kohlrausch-type. The jumps occurring when the photon flux is switched on and off are due to heating/cooling effects which are estimated to be ~ 1.5K.

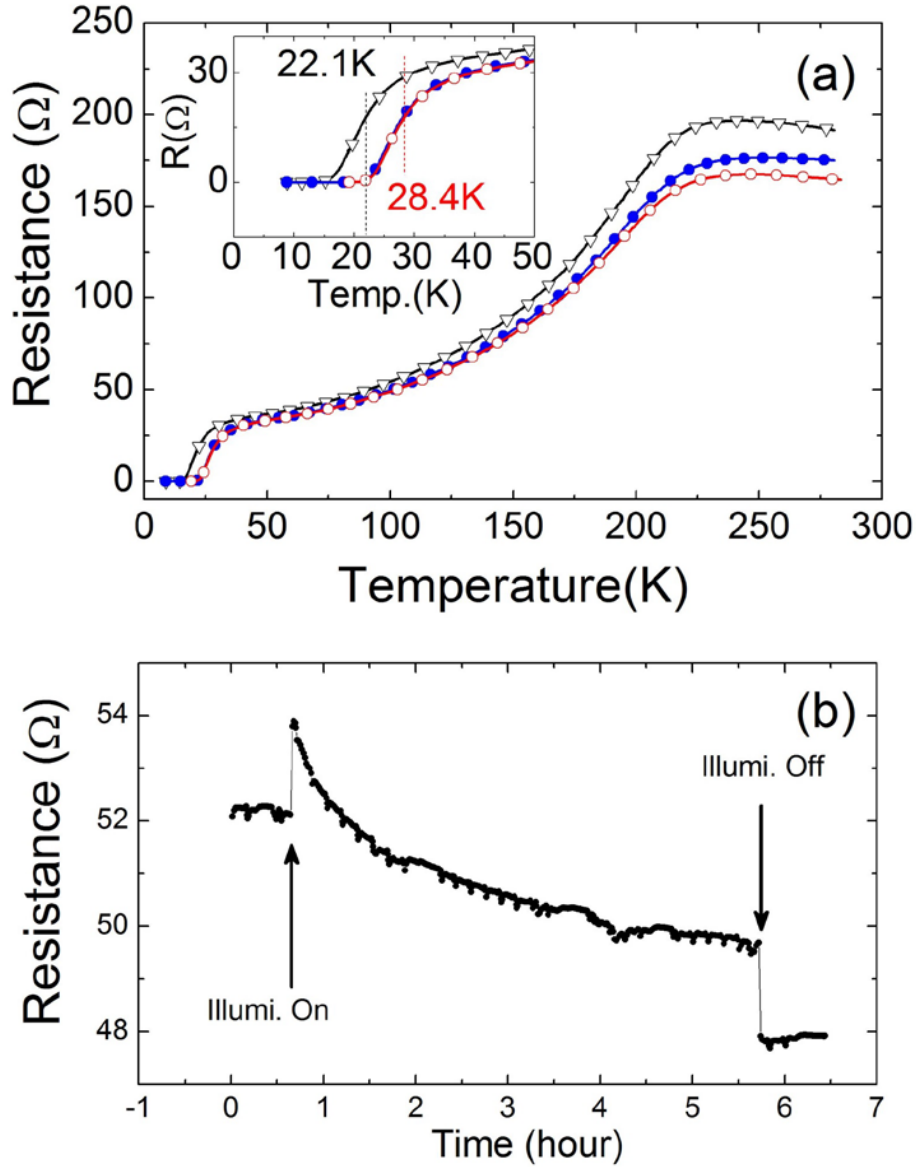

Fig. 2. (a) Temperature dependence of the resistance of a oxygen depleted SL sample (x= 6.5). Resistance in the dark (open triangles), after illumination (filled circles) measured in the dark, and after illumination measured with photon exposure (open circles ). The inset is the magnification of the low temperature region. (b) Resistance measurement during the illumination as a function of time.

In the same way as described above we measured a set of oxygen-depleted YBCO single layer films as well as superlattices. The results are summarized in Fig 3 in a ΔTc vs. Tc plot together with data for single layer YBCO films taken from Osquiguil.[17] The error bars in Fig. 3 are drawn from the onset of superconductivity to the zero resistance values in the corresponding R (T) curves. Fig.3 clearly shows a PPC effect in the superlattices and its increase with decreasing $T_C$ that coincides within the error bars with that of single layer YBCO films.

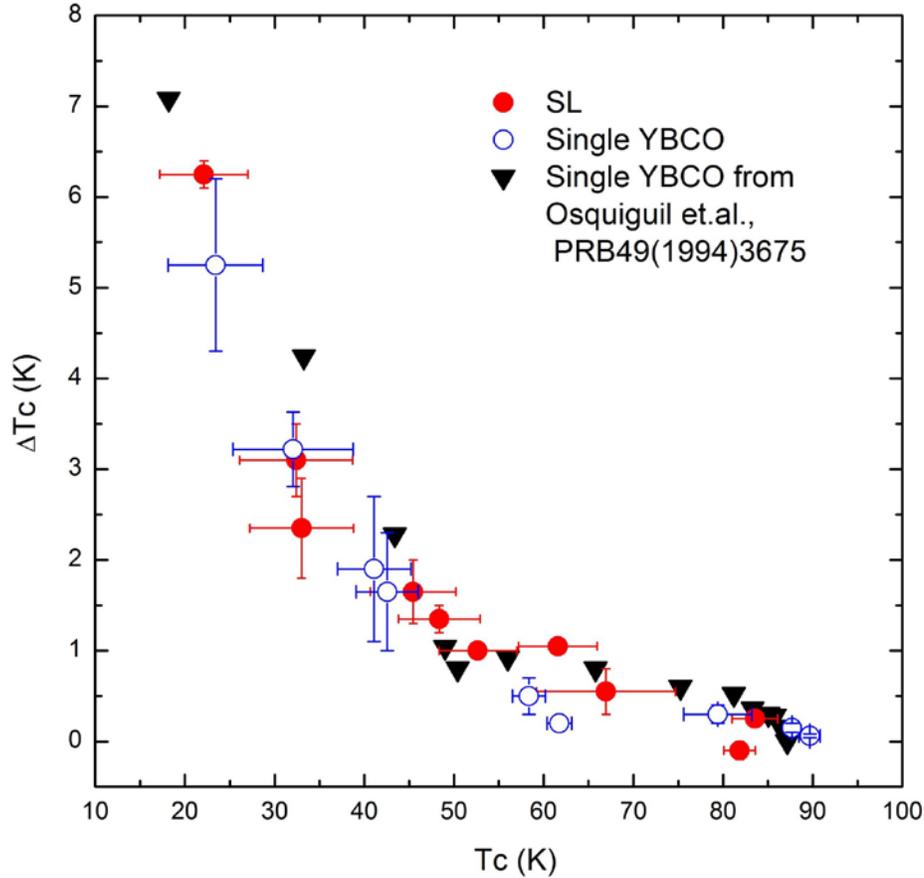

Fig. 3 Photoinduced persistent enhancement of $T_c$, in a $\Delta T_c$ vs $T_c$ plot. The superlattice data ( filled circles ) are compared with single layer YBCO films of this work (open circles) and literature values (filled triangles)

To discriminate the $T_C$ reduction due to oxygen depletion and the subsequent increase by photon illumination from that due to superlattice formation, we prepared a set of fully oxygenated SL films with different YBCO layer thicknesses $[LCMO_{20}/YBCO_x]_{10}$ ranging from x= 8, 4.5 and 3.5 nm, respectively. The resulting R (T) data are represented in Fig. 4. The resistances in the dark show the large reduction of the midpoint of the transition to superconductivity as the YBCO layer becomes thinner, i.e. $T_c$ = 70, 2 K, 58 K and 35.3 K for x=8, 4.5 and 3.5, respectively. The data clearly show the absence of a photoinduced resistance change. The reduction of $T_C$ for films with decreasing YBCO layer thickness is due to the electronic reconstruction at the YBCO/LCMO interface as discussed previously.[5-7]

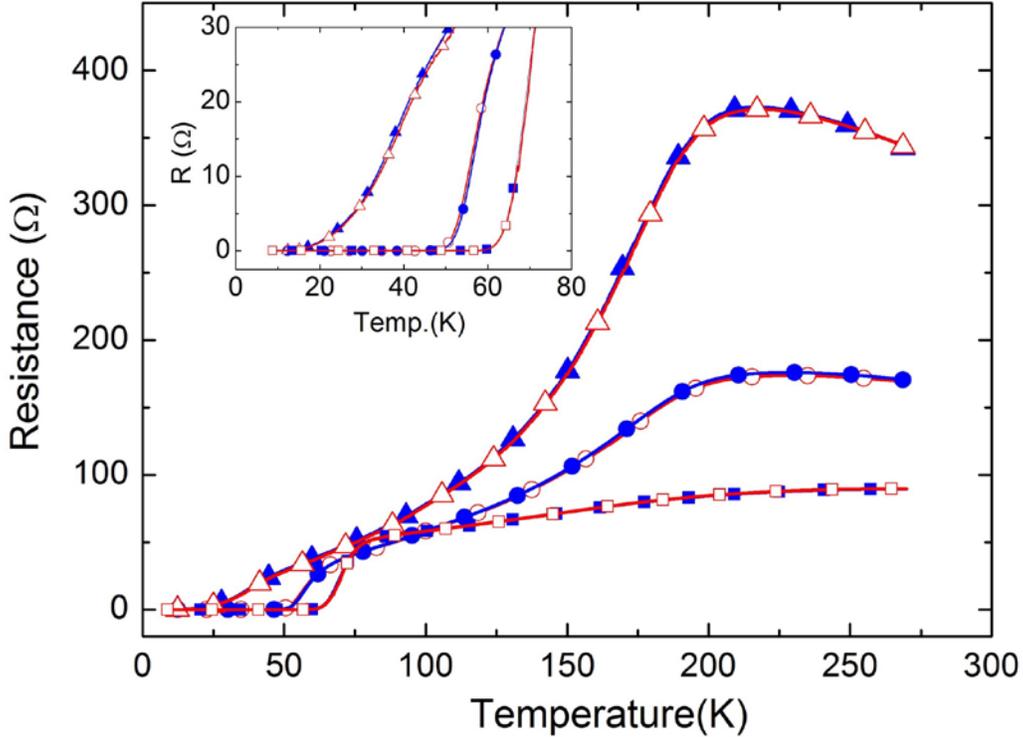

Fig. 4. Temperature dependencies on the resistance of SL samples [LCMO$_{20}$/YBCO$_x$]$_{10}$ with x=8, 4.5 and 3.5nm, respectively. The open and filled symbols refer to the data before and after the illumination. The inset shows the magnification of the low temperature region.

Comparing our results for oxygen-depleted YBCO/LCMO superlattices grown at 720$^0$C by PLD and those of Peña et al.[20] for YBCO/LCMO bilayer films sputtered at 900$^0$C, several differences are apparent. First, the sputtered bilayers show no PPC effect, while sputtered single layer YBCO films do. Second, the $T_c$ enhancement amounts to up to 23 K in the sputtered bilayers; these values are about 6 times larger than those measured in sputtered single layer films of identical degree of oxygenation.[20] In contrast, PLD-grown superlattices show a persistent photoconductivity with a $\Delta T_C$ up to ~6 K depending on the degree of oxygenation; the $T_C$ dependence of the PPC effect matches nicely that of single layer YBCO films.

Following the arguments given by Peña et al.[20] the transient photoconductivity and absent PPC observed in sputtered bilayer YBCO/LCMO structures are due to interfacial and/or electron transfer effects. The enhancement of $T_C$ must arise from the YBCO layer where a surplus of holes appears during photon exposure. An increased relaxation compared to single layer YBCO films is induced by the LCMO layer. The similarity of the time scales of the relaxation time for the bilayer and single layer LCMO in Peña's experiments suggest that charge transfer must be the driving force for the transient photoconductivity and is controlled by the LCMO layer. On the other hand, this implies that the absence of a transient photoconductivity and the presence of a persistent photoconductivity in the PLD-grown superlattices, together with the quantitative agreement of $\Delta T_C$ for single layers and superlattices is caused by the lack of or drastically reduced charge transfer. The reason for the difference between sputtered and PLD grown films may rest in details of the microstructure of the films especially the structure of the interface. Careful transition electron microscopy (TEM) analysis of PLD grown YBCO-LCMO-YBCO trilayer systems by Zhang et al.[26] reveal that the layers are epitaxial, atomically smooth, and uniform across the entire specimen captured by the TEM experiments. The atomic stacking sequence at the interface of the LCMO layer grown on YBCO is different from that of the YBCO layer grown on LCMO. Both sequences, however, are different from the configuration previously identified in YBCO/LCMO superlattices prepared by sputtering using conditions identical to those used by Peña et al.[27]

In conclusion, PLD-grown oxygen depleted YBCO/LCMO superlattices show a persistent photoconductivity scaling nicely with those of single layer YBCO thin films. This is interpreted as a consequence of the lack of massive charge transfer from YBCO to LCMO and an enhanced recombination there. The origin is seen in details of the microstructure of the films and that of the interfaces which in turn sensitively depend on the deposition technique and the fine tuning of the deposition process. Consequently, at the current stage of even advanced complex oxide deposition technologies it is difficult to generalize results without in depth knowledge of the structure of interfaces at the atomistic level and the role of process-induced defects in films and interfaces.

**Acknowledgments**

The authors thank Cameron Hughes for carefully reading the manuscript and G. Christiani for experimental advice. Parts of the work has been supported by the DFG (contract HA 1415/15-1)